\documentstyle[epsfig,12pt,mathptmx]{article}
\begin{document}
{\large\bf New nuclear three-body clusters $\,\phi$\textit{NN}}\\
[4mm]
%
{V. B. Belyaev$^a$, W. Sandhas$^b$, and I. I. Shlyk$^a$}\\
{\small \em $^a$Joint Institute for Nuclear Research, BLTP, Dubna,
141980, Russia\\
$^b$Physikalisches Institut, Universitat Bonn,
D-53115 Bonn, Germany}
\vspace{6mm}

\noindent{\small\bf Abstract.}
 {\small\  Binding energies of three-body systems of
the type $\,\phi$+2\textit{N} are estimated. Due to the strong
attraction between $\phi$-meson and  nucleon, suggested in
different approaches, bound states can appear in systems like
$\phi$+\textit{np} (singlet and triplet) and $\phi$+\textit{pp}.
This indicates the principal possibility of the formation of new
nuclear clusters. }
\vspace{6mm}


\noindent The interaction of mesons containing heavy quarks with
nucleons may shed some light on two phenomena, which have not yet
been studied sufficiently.

One of these problems, considered already in [1-3], deals
with the interaction of charmonium mesons with nucleons. It is obvious,
that the main contribution to this interaction will originate not from the
usual ~$\pi,\,\rho,\,\omega,\,\sigma$ - meson exchange, but from
gluon exchange which means that a new type of short- range forces
enters the game.

 Important is the fact, that  this initial degrees of freedom of QCD manifests itself at low energies of the
 interacting particles, i.e., of charmonium mesons and nucleons. The second
 phenomenon, appearing especially in the ~$\phi\,-\,N$
 system, is the considerable attraction predicted for these
 particles. Employing some phenomenological model and assuming a
 dominant role of the ~$s\overline{s}$ configuration
 in~$\phi$- meson structure, the authors of ~\cite{Gao}, following ideas
 of ~\cite{Brodsky}, suggested an attractive Yukawa type potential
 for the ~$\phi\,-\,N$ interaction
\begin{equation}
\label{Yukawa} V_{\phi\,n}(\,r)\;=\;-\,\alpha\,e^{-\,\mu\,r}/\,r,
\end{equation}
where $\alpha\;=\;1.25$ and $\mu\;=\;600$~MeV which supports a
bound state in this system of about 9.3 MeV binding energy.

 In a rather different approach, based on the quark model,
 the authors of ~\cite{Huang} have shown, by applying Resonating Groupt techniques,
 that quite strong attraction may appear in the~$\phi\,-\,N$ system.
In both models the~$\phi\,-\,N$ interaction is of very short range
~$r_{\,0}\;\sim\,\frac{\,1}{\,m}$ , where $m\; \sim\;600$ MeV. All
these observations open an interesting possibility to form new
 few- body clusters,
 like~$\phi\,+\,2n,\,\phi\,+\,2p,\,\phi\,+\,3n$.

In what follows we present calculations of binding energies for
the systems~$\phi\,n\,p$ ($^1 S_{\,0}$ for $n\,p$) , $\phi\,n\,p$
($^3 S_{\,1}$ for $n\,p$)  and $\phi\,p\,p$ based on the folding
approximation.

The variational method with a trial function
\begin{equation}
\label{psi_phi_n}
\Psi_{\phi\,n}(r)\;\sim\;e^{\,-\,\alpha_{\,1}\,r}\,+\,c\,e^{\,-\,\alpha_{\,2}\,r},
\end{equation}
where~$r$ is the distance between the $\phi$- meson and a neutron
while $\alpha_{\,1},\,\alpha_{\,2},\,c$ are variational
parameters, gives for $\phi\,-\,n$ a binding energy of about 9.3
MeV which is quite close to the exact numerical solution.
 Motivated by this observation we use the same trial function with the same  variational
 parameters in  calculating the binding energy of the~$\phi\,+\,2n$ system by means of the
 folding method.

 The Schr\"{o}dinger equation in Jacobi coordinates reads
\begin{equation}
\label{Schrodinger equation}
\left\{-\,\frac{\,\hbar^{\,2}}{\,2\,\mu_{\,1}}\,\Delta_{\,\overline{\,r}}\,-
\,\frac{\,\hbar^{\,2}}{\,2\,\mu_{\,2}}\,\Delta_{\,\overline{\,R}}\,+\,V\right\}\Psi\;=\;E\,\Psi
\end{equation}
where~$\mu_{\,1}$ is the reduced mass of the $\phi$- meson and a
neutron, while $\mu_{\,2}$ denotes the reduced mass of the
~$\phi\,n$ cluster and a neutron,
\\
$$V\;=\;V_{\phi\,n_{\,1}}(r)\,+\,V_{\phi\,n_{\,2}}(\overline{r},\overline{R})\,+
\,V_{n_{\,1}\,n_{\,2}}(\overline{r},\overline{R}).$$

For the sake of simplicity, we consider Malfliet-Tjon
nucleon-nucleon singlet (\,I) and triplet (\,III)
potentials~\cite{Tjon}
\begin{equation}
\label{Tjon}
V(\,r)\;=\;-\lambda_{\,A}\,\frac{e^{-\mu_A\,r}}{\,r}\,+\,\lambda_{\,R}\,\frac{e^{-\mu_R\,r}}{\,r}
\end{equation}
where $\lambda_{\,A}\;=\;2.64$, $\mu_A\;=\;1.55$ fm$^{\,-1}$,
$\lambda_{\,R}\;=\;7.39$, $\mu_R\;=\;3.11$ fm$^{\,-1}$ for singlet
potential (\,I) and $\lambda_{\,A}\;=\;3.22$, $\mu_A\;=\;1.55$
fm$^{\,-1}$, $\lambda_{\,R}\;=\;7.39$, $\mu_R\;=\;3.11$
fm$^{\,-1}$ for triplet potential (\,III).\\
 Following the idea of the folding method, we
represent the full function as
$$\Psi\;=\;\Psi_{\phi\,n}(r)\,\chi(\overline{R})
$$
where $\Psi_{\phi\,n}(r)$ is defined in (\ref{psi_phi_n}) and
$\chi(\overline{R})$ is an unknown spectator-function. Using this
representation and the equation (\ref{Schrodinger equation}),we
easily find
\begin{equation}
\label{folding}
-\,\frac{\,\hbar^{\,2}}{\,2\,\mu_{\,2}}\,\Delta_{\,\overline{\,R}}\,\chi(\overline{R})\,+\,\widetilde{V}(R)\,
\chi(\overline{R})\;=\;(E-E_{\,0})\,\chi(\overline{R}).
\end{equation}

 Here~$E_{\,0}\;=\;-\,9.3$ MeV and
$$\widetilde{V}(R)\;=\;\frac{\langle \Psi_{\phi_n}|V_{\phi\,n_{\,2}}|\Psi_{\phi_n}\rangle \,+
\,\langle \Psi_{\phi_n}|V_{n_{\,1}\,n_{\,2}}|\Psi_{\phi_n}\rangle
}{\langle \Psi_{\phi_n}|\Psi_{\phi_n}\rangle
}
$$
is the effective potential, which can be calculated analytically.

 When calculating~$\widetilde{V}(R)$ and solving~(\ref{folding}) we
find for $\phi\,n\,p$ ($^1 S_{\,0}$ for $n\,p$) a binding energy
of about 10.1~MeV. The same procedure yields for $\phi\,n\,p$ ($^3
S_{\,1}$ for $n\,p$)  17.7~MeV. Now we can estimate the binding of
$\phi\,p\,p$ simply by adding to the singlet $n\,p$ interaction
the Coulomb potential. This results in a binding near 9.6 MeV.

 Consequently, a bound $\phi\,+\,2N$ system may exist which
would represent a new kind of nuclear clusters.

Let us emphasize some peculiarity of such a system. By comparing
(\ref{Yukawa}) and (\ref{Tjon}) and the parameters of these
potentials, the repulsive part of the Malfliet- Tjon potential has
a wider range than the attraction area of the $\phi- n$ potential.
Roughly speaking this means that only the tail of the attractive
$\phi- n$ potential is responsible for the binding of the
 whole system. In other words, the binding of the system is defined
 by a delicate balance of a narrow~$\phi- n$ attraction and the
 position and the size of a repulsive part of the nucleon- nucleon
 interaction. In this way it is apparent that the properties of
 such a system are very sensitive to the short- range behavior
 of the $NN$- potentials.

\end{document}